\def\slash#1{\setbox0=\hbox{$#1$}#1\hskip-\wd0\dimen0=5pt\advance
\dimen0 by-\ht0\advance\dimen0 by\dp0\lower0.5\dimen0\hbox
to\wd0{\hss\sl/\/\hss}}
\def\Black{}
\def\Blue{}
\def\Brown{}
\documentstyle[11pt,mo2,epsfig]{article}
\newcommand{\be}{\begin{equation}}
\newcommand{\ee}{\end{equation}}

\begin{document}
\begin{titlepage}
\null
\begin{center}
\Large\bf \Brown A Quark model for Heavy Mesons:\\
Strong Decays \Black
\end{center}
\vspace{0.5cm}
\begin{center}
\begin{large}
A. D. Polosa\\
\end{large}
\vspace{0.3cm}
Dipartimento di Fisica, Universit\`a di Bari and INFN Bari,\\
via Amendola 173, I-70126 Bari, Italia
\end{center}
\vspace{0.5cm}
\begin{center}
\begin{large}
\Brown
{\bf Abstract}\\[0.5cm]\Black
\end{large}
\parbox{14cm}{I shortly review the application of a recently introduced
constituent quark-meson model to the determination of some strong
coupling constants governing the decays of charm heavy mesons with
one pion in the final state. I present the model directly through
its rules for computing amplitudes. For a detailed explanation of
the model, the reader is referred to the original papers.
}
\\
\vspace{1.0cm}
\noindent
\Blue
\parbox{14cm}{
}
\Black
\end{center}
\vspace{3.0cm}
\begin{center}
{\it To appear in the Proceedings of XIth Rencontres de Blois\\
Blois, France, June 27-Jul 3, 1999}
\end{center}
\vfil
\noindent
\Brown
BARI-TH/99-360\\
Sept 1999
\Black
\end{titlepage}



\bibliographystyle{unsrt}

\def\Journal#1#2#3#4{{#1} {\bf #2}, #3 (#4)}

\def\NCA{\em Nuovo Cimento}
\def\NIM{\em Nucl. Instrum. Methods}
\def\NIMA{{\em Nucl. Instrum. Methods} A}
\def\NPB{{\em Nucl. Phys.} B}
\def\PLB{{\em Phys. Lett.}  B}
\def\PRL{\em Phys. Rev. Lett.}
\def\PRD{{\em Phys. Rev.} D}
\def\ZPC{{\em Z. Phys.} C}

\def\st{\scriptstyle}
\def\sst{\scriptscriptstyle}
\def\mco{\multicolumn}
\def\epp{\epsilon^{\prime}}
\def\vep{\varepsilon}
\def\ra{\rightarrow}
\def\ppg{\pi^+\pi^-\gamma}
\def\vp{{\bf p}}
\def\ko{K^0}
\def\kb{\bar{K^0}}
\def\al{\alpha}
\def\ab{\bar{\alpha}}
\def\be{\begin{equation}}
\def\ee{\end{equation}}
\def\bea{\begin{eqnarray}}
\def\eea{\end{eqnarray}}
\def\CPbar{\hbox{{\rm CP}\hskip-1.80em{/}}}
\vspace*{4cm}
\title{A QUARK MODEL FOR HEAVY MESONS:\\ STRONG DECAYS}

\author{ A.D. POLOSA }

\address{Dipartimento di Fisica, Universit\`a di Bari and INFN Bari, Via Amendola 173,\\
I-70126 Bari, Italy}

\maketitle\abstracts{
I shortly review the application of a recently introduced
constituent quark-meson model to the determination of some strong
coupling constants governing the decays of charm heavy mesons with
one pion in the final state. I present the model directly through
its rules for computing amplitudes. For a detailed explanation of
the model, the reader is referred to the original papers.}

\section{Introduction}

A Constituent-Quark-Meson model, CQM, has been recently discussed
in a number of papers \cite{art1,art3,aldo} stimulated by the
original proposals of Ebert et al. \cite{ebert}. The main feature
of this model is the emergence of effective (heavy meson)-(heavy
quark)-(light quark) vertices. The effective Lagrangian predicting
them, comes out from the bosonization of a primary
Nambu-Jona-Lasinio (NJL) interaction involving light and heavy
quark fields \cite{ebert}. This heavy-light sector Lagrangian,
${\cal L}^{hl}$, incorporates heavy-flavour symmetries.

The heavy and light quarks are free in the meson because the model
is not confining, however, unphysical thresholds for meson decays
into real quark-antiquark pairs are avoided through the
introduction of an infrared cutoff $\mu$ which prevents to explore
energy regions below $\Lambda_{\rm QCD}$.

The constituent light quark mass, $m=300$ MeV, is dynamically
generated by a NJL-gap equation and acts as the order parameter
characterizing the transition between the broken and unbroken
chiral symmetry phases. The light sector Lagrangian, ${\cal
L}^{ll}$, incorporates chiral symmetry and its spontaneous
breakdown and resembles the Manohar-Georgi \cite{manogeo} effective
Lagrangian, except for  the absence of gluons and for a different
structure of the light quark fields, remnant of the underlying NJL
interaction.

\begin{figure}[t!]
\begin{center}
\epsfig{bbllx=0.5cm,bblly=16cm,bburx=20cm,bbury=23cm,height=5truecm,
        figure=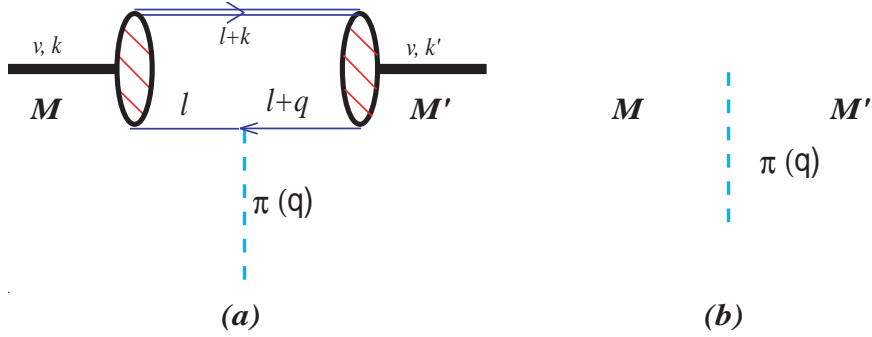}
\caption{\label{fig:figura}
\footnotesize
          The CQM loop diagram is depicted in (a) while, in (b), is represented the same
          process described at the level of a mesonic effective Lagrangian, i.e., a
          Lagrangian in which the fundamental fields are meson fields. }
\end{center}
\end{figure}

\section{Strong decays}

In this paper I describe the application of CQM model to the
evaluation of the strong coupling constants governing the decays
$H\to H\pi$, $S\to H\pi$, $S\to S\pi$, $T\to H\pi$, $T\to S\pi$,
where $H$, $S$ and $T$ are the superfields \cite{falco}
representing respectively, the lowest negative parity spin doublet,
$(0^-,1^-)$, and the higher spin positive parity doublets,
$(0^+,1^+)$ and $(1^+,2^+)$, predicted by the Heavy Quark Effective
Theory (HQET).

Such processes involve amplitudes that can be calculated
in the CQM model through simple loop diagrams, such that in Fig.
1(a), in which the pion is coupled to the light quark line,
according to the prescription in ${\cal L}^{ll}$, and mesons are
attached to the ends by means of the effective vertices predicted
in  ${\cal L}^{hl}$.

The explicit calculation of such a diagram proceeds as an usual
loop calculation in which the Feynman rules, derived by CQM, are
used. With reference to Fig. 1(a), the loop integral is written as:
\begin{equation}
(-1)i^3 i^3 \sqrt{Z_{\cal M}Z_{\cal M^\prime}m_{\cal M}m_{\cal
M^\prime}} \frac{3}{16\pi^4}\int^{\rm reg} d^4\ell
\frac{{\rm Tr}\left[(\gamma\cdot \ell+m)\left(-\frac{q^\mu}{f_\pi}\gamma_\mu \gamma_5
\right)(\gamma\cdot (\ell+q)+m){\cal M^\prime}(v){\cal
M}(v)\right]}{(\ell^2-m^2)((\ell+q)^2-m^2)(v\cdot \ell+\Delta_{\cal
M})},
\end{equation}
where ${\cal M}$ and ${\cal M^\prime}$ are respectively the
incoming and outcoming meson fields of the $H$, $S$ and $T$ type
\cite{falco}. $Z_{\cal M}$ and $Z_{\cal M^\prime}$, are the heavy
meson field renormalization constants \cite{art1}. $m_{\cal M}$ and
$m_{\cal M^\prime}$, are the masses of the degenerate heavy meson
doublets, ${\cal M}$ and ${\cal M^\prime}$, introduced according to
the normalizations of the annihilation operators in $H$, $S$ and
$T$ \cite{art1}. $(-1)$ comes from the Furry theorem, $i^3$ from
propagators and the other $i^3$ factor from the three effective
vertices $q\pi q$, $q{\cal M}Q$ and $Q{\cal M^\prime}q$, where $q$
and $Q$ denote respectively the light and heavy quark fields. The
term $\left(-\frac{q^\mu}{f_\pi}\gamma_\mu
\gamma_5\right)$ is responsible for the derivative coupling of {\it
one} pion to the light quark line. The integral should be
calculated using the Schwinger proper time regularization method
which allows to exponentiate the light quark propagators and to cut
off the momenta running in the loop, according to the prescription:
\begin{equation}
\int d^4 \ell_E \frac{1}{\ell_E^2+m^2}\to\int d^4\ell_E \int_{1/\Lambda^2}^{1/\mu^2}ds
e^{-s(\ell_E^2+m^2)},
\end{equation}
where, $\Lambda
\simeq \Lambda_\chi\simeq4\pi f_\pi\simeq 1$ GeV, is the ultraviolet
cutoff \cite{manogeo} and $\mu=300$ MeV.

It is possible to perform a calculation {\it beyond the soft pion
limit} \cite{art3}, retaining $q$ in (1), being
$q^\mu=(q_\pi,0,0,q_\pi)$ the pion momentum in the chiral limit and
$q_\pi=\Delta_{\cal M}-\Delta_{\cal M^\prime}\neq 0$. $\Delta_{\cal
M}$ and $\Delta_{\cal M^\prime}$ are the mass differences $m_{\cal
M}-m_Q$, $m_{\cal M^\prime}-m_Q$ and $m_Q$ is the mass of the heavy
quark involved.

The result of the CQM loop calculation must be compared with the
explicit expression one founds for the matrix element:
\begin{equation}
\langle {\cal M^\prime} \pi|i{\cal L}_{\rm meson}|{\cal M}\rangle,
\end{equation}
where ${\cal L}_{\rm meson}$ is the appropriate interaction term
describing the transition ${\cal M}\to {\cal M^\prime}$, as in Fig.
1(b), at the level of heavy meson chiral Lagrangian \cite{wise}.

This comparison allows to extract the strong coupling constants
introduced at the level of ${\cal L}_{\rm meson}$, e.g., $g$
defined by ${\cal L}_{\rm meson}=ig{\rm Tr}\left[\bar{H}
H\left(-\frac{q^\mu}{f_\pi}\gamma_\mu
\gamma_5\right)\right]$. With this approach, the results reported
in Table 1 have been obtained. In Table 1 we refer to charm states
because they are at the moment object of experimental search. CQM
results seem encouraging if compared with other theoretical
calculations. For example, the determination of the strong coupling
in the process $S\to H\pi$ turns out to be very close to the QCD
sum rule determination \cite{art3}. The same happens for $g$
\cite{art1}.

\begin{table}[t]
\caption{Strong coupling constants, with one pion in the final
state, calculated through the CQM model. It is obvious to assume
the soft pion limit $q_\pi \to 0$ (s.p.l.) for the processes $H\to
H\pi$ and $S\to S \pi$. Some recent CLEO data
give a mass $m_S$, for the $S$ multiplet, very close to $m_T$.
Therefore the (s.p.l.) is applied also in the calculation of the
process $T\to S \pi$. The theoretical error here reported is
calculated letting the parameter $\Delta_H$ vary in the range
$0.3,0.4,0.5$ GeV ($\Delta_S$ and $\Delta_T$ vary accordingly). $k$
and $\tilde{h}$ are published here for the first time.}
\vspace{0.4cm}
\begin{center}
\begin{tabular}{|c|c|l|}
\hline
${\rm Decay\;channel}$ & ${\rm Coupling\;constant}$ & ${\rm
CQM\;result}$
\\ \hline
$D^*\to D\pi$ & $g$ & $0.46\pm 0.04\;{\rm (s.p.l.)}$
\\ \hline
$D_0\to D\pi$ & $h$ & $-0.56\pm 0.11$
\\ \hline
$D_1^*\to D^*\pi$ & &
\\
$D_2^*\to D\pi$ & $h^\prime$ & $0.65^{+0.45}_{-0.30}$
\\
$D_2^*\to D^*\pi$ & &
\\ \hline
$D_0\to D_1\pi$ & $k$ & $-0.13\pm0.05\;{\rm (s.p.l.)}$
\\ \hline
$D_1^*\to D_0\pi$ & &
\\
$D_1^*\to D_1^{*\prime}\pi$ & $\tilde{h}$ &
$0.91^{+0.5}_{-0.3}\;{\rm (s.p.l.)}$
\\
$D_2^*\to D_1^{*\prime}\pi$ & &
\\ \hline
\end{tabular}
\end{center}
\end{table}
\section*{Acknowledgments}
I would like to thank A. Deandrea, R. Gatto and G. Nardulli for
discussions.

\end{document}